\documentclass[prl,twocolumn]{revtex4}
\usepackage[english]{babel}
\usepackage{amssymb}  
\usepackage{amsmath}
\usepackage{graphicx}
\usepackage{xcolor}
\usepackage[pdftex,colorlinks=true,linkcolor=blue,citecolor=blue,urlcolor=blue,filecolor=blue,breaklinks=true]{hyperref}  %clickable links in refs
\usepackage{url}
\definecolor{dark-blue}{rgb}{0.15,0.15,0.4}

\begin{document}

\title{Reply to: `Comment on: ``The chiral phase transition in charge ordered {\it 1T}-TiSe$_2$" '}

\author{Stephan Rosenkranz$^1$}
\author{Ray Osborn$^1$}
\author{Jasper  van Wezel$^{2}$}
\email{vanwezel@uva.nl}
\affiliation{
$^1$ Materials Science Division, Argonne National Laboratory, Argonne, IL 60439, USA.\\
$^2$ Institute for Theoretical Physics Amsterdam, University of Amsterdam, Science Park 904, 1098 XH Amsterdam, The Netherlands.
}

\begin{abstract}
We offer a reply to the recently posted comment on our earlier work concerning the chiral phase transition in charge ordered {\it 1T}-TiSe$_2$.
\end{abstract}

\maketitle

%%%%%%%
A comment on our description~\cite{us} of the chiral phase transition in charge ordered {\it 1T}-TiSe$_2$ was recently posted on the arxiv~\cite{them}. The comment offers an alternative interpretation that addresses part of the x-ray data in our original work, but is incompatible both with the behavior of other physical observables that we presented~\cite{us}, and with recent direct observations of the broken inversion symmetry in  {\it 1T}-TiSe$_2$~\cite{MIT}.

It is a well-known fact of scientific life that isolated experimental observations
are open to interpretation. It is therefore not surprising there may exist
multiple ways of interpreting the one figure reproduced by the authors
of this comment. The fact that alternative interpretations of a partial
data set exist, is not in and of itself an indication that the original interpretation is flawed.

In this regard, the authors of the comment ignore the fact that the
figure they reproduce represents only part of the data presented in Ref.~\citenum{us}.
The alternative interpretation they give for the one isolated figure is in fact
inconsistent with the rest of the data presented in the original article~\cite{us}. That is, assuming the alternative
interpretation that there is only a single transition in {\it 1T}-TiSe$_2$,
it is not possible to explain the two kinks in the specific heat, nor the
two anomalies in the resistance, both measured on the same sample as that used
for the x-ray experiment, and both occurring at the same second transition temperature $7$\,K below the main CDW transition temperature
$T_{CDW}$.

Our original interpretation~\cite{us} was based on three sets of
different physical properties (x-ray data, specific heat, electrical resistance), providing a consistent explanation for that entire set of
observations. In contrast, the alternative interpretation presented in the comment~\cite{them} does not give a consistent explanation of the entire data set.

Finally, we would like to point out that direct evidence for the breakdown of inversion symmetry  in {\it 1T}-TiSe$_2$, consistent with the presence of a chiral charge density wave, was recently reported on the basis of circularly polarised photogalvanic effect measurements~\cite{MIT}.

\bibliographystyle{apsrev4-1}
\bibliography{reply}

%\begin{thebibliograhpy}
%\end{thebibliography}

\end{document}